\documentclass{webofc}
\usepackage[varg]{txfonts}   % Web of Conferences font
\usepackage{hyperref}
\usepackage{url}
\hypersetup{colorlinks=true,citecolor=blue,urlcolor=blue,linkcolor=blue}
\usepackage{amsmath,amssymb,bm,bbm}
\usepackage{graphicx}
\usepackage{multirow}
\usepackage{color}
\usepackage[normalem]{ulem}
\usepackage{tikz}

\usepackage{float}
\restylefloat{figure}

\usepackage{mathrsfs}
\usepackage[singlelinecheck=false]{caption}

\newcommand{\ket}[1]{ | #1 \rangle }

\newcommand{\overlap}[2]{\langle #1 | #2 \rangle}
\newcommand{\elmx}[3]{\langle #1 | #2 | #3 \rangle}

\begin{document}
\title{Discrete non-orthogonal shell model for nuclear structure: towards heavy elements}

\author{\firstname{Duy Duc} \lastname{Dao}\inst{1}\fnsep\thanks{\email{duc.dao@iphc.cnrs.fr}} \and
\firstname{Fr\'ed\'eric} \lastname{Nowacki}\inst{1}\fnsep\thanks{\email{frederic.nowacki@iphc.cnrs.fr}}
}

\institute{Université de Strasbourg, CNRS, IPHC UMR7178, 23 rue du Loess, F-67000 Strasbourg}

\abstract{%200
We present recent developments of the Discrete Non-Orthogonal Shell Model (DNO-SM) for nuclear structure studies far from stability. Exact shell-model solutions are obtained for typical open-shell light sd and pf nuclei using non-orthogonal Slater determinants consistently derived from the variation after projection approach. The latter represents a powerful method to include correlations from particle-hole excitations. Applications to proton-rich nuclei at the $N\sim Z$ line show the important role of these correlations to probe the structure transition in the $^{84,86}$Mo isotopes. We finally present a first complete description of low-lying spectroscopy in the superheavy $^{254}$No, reproducing excellently various band structures and isomers in this challenging nucleus. 
}
\maketitle
\section{Introduction}
\label{introduction}

The nuclear Shell Model has become now one of the major assets to nuclear structure studies. Its success is due to impressive computational advances over many years of developments~~\cite{Caurier2005,BigStickJOHNSON20132761,NuShellXBROWN2014,NOWACKI2021} along with a thorough understanding of the effective nuclear interaction~\cite{Caurier2005,NOWACKI2021}. One of the challenges within the shell-model framework is to efficiently tackle the secular diagonalisation problem. In that perspective, variational methods based on symmetry-breaking mean-field states have been studied since the 1980s to address this question~\cite{Schmid1984A_PhysRevC.29.291,Schmid2004}. An important issue in this line of approach, which still remains open today, is the incorporation of nuclear ground-state correlations to recover the shell-model ground-state solution. In relation to this question, our recent Discrete Non-Orthogonal Shell Model (DNO-SM)~\cite{DNO2022_PhysRevC.105.054314,DNOVAP2025} has been inspired notably from the Broeckhove-Deumens theorem~\cite{Deumens1979} which stipulates the existence of a discrete set of non-orthogonal Slater determinant states spanning the shell-model valence space. In the following, we will first present benchmark results which provide a numerical evidence of this theorem. In particular, using non-orthogonal Slater states as the variational wave function ansatz, we study the description of pairing correlations in a textbook shell-model example of $^{48}$Cr. Finally, recent applications of the DNO-SM to proton-rich Molybdenum isotopes~\cite{Recchia2025} and a first shell-model description of the superheavy $^{254}$No~\cite{No254ARXIV2025} will be discussed.

\section{Discrete non-orthogonal shell model}
\label{sect1}
\subsection{Projection-after-Variation versus Variation-after-Projection approaches}
In the variational framework, the nuclear wave function is generally represented as a superposition of non-orthogonal intrinsic states $\ket{\phi}$ restored to good angular momentum $J$ and parity $\pi$ for example,
\begin{equation}
    \ket{\psi^{J,\pi}_\alpha} = \sum_{K} C^{J\pi}_\alpha \mathcal P^{J}_{MK} \mathcal P^\pi \ket{\phi}.
\end{equation}
where $\mathcal P^{J}_{MK}$ is the Peierls-Yoccoz ansatz for angular momentum projection~\cite{RingSchuck1980} and $\mathcal P^\pi$ the parity projector. For a given effective Hamiltonian $\hat H$, the projected energy can be calculated as the expectation value
\begin{equation}
    E_\alpha^{J\pi}[\phi] = \frac{\elmx{\psi^{J,\pi}_\alpha}{\hat H}{\psi^{J,\pi}_\alpha}}{\overlap{\psi^{J,\pi}_\alpha}{\psi^{J,\pi}_\alpha}}, 
    \label{Eproj}
\end{equation}
which is an energy functional of the intrinsic state $\ket\phi$. One then distinguishes between a projection after variation (PAV) and a variation after projection (VAP) approach depending on whether the symmetry restoration is taken into account in the energy variation~\cite{RingSchuck1980}. In principle, the VAP approach is the most consistent with a proper treatment of broken symmetries by the minimization of the projected energy~\eqref{Eproj}. However, for this reason, it is also generally more complicated. Following~\cite{Schmid2004}, starting from a initial state $\ket{\phi_0}$, one can look for a better solution $\ket{\phi} = \mathcal N_0\displaystyle e^{\sum_{ij}Z_{ij}}a^\dagger_i a_j\ket{\phi_0}$ using Thouless parametrization of non-orthogonal states. The first step is to tackle the Hill-Wheeler-Griffin equation
\begin{equation}
    H^{J\pi} C^{J\pi}_\alpha = E_\alpha^{J\pi} N^{J\pi} C^{J\pi}_\alpha, \:\text{with }
    H^{J\pi}_{KK'} = \elmx{\phi}{H\mathcal P^J_{KK'}}{\phi}, \:
    N^{J\pi}_{KK'} = \elmx{\phi}{\mathcal P^J_{KK'}}{\phi}.
    \label{HWG}
\end{equation}
From the resulting eigensolution, the second step will be the evaluation of the energy variation $\frac{\delta}{\delta\phi}E_\alpha^{J\pi}[\phi] = \frac{\partial}{\partial Z^*}E_\alpha^{J\pi}$ which yields a descent direction for the minimization of the projected energy~\eqref{Eproj}. A better approximation of $\ket\phi$ can then be obtained and one proceeds again the resolution of the Hill-Wheeler-Griffin equation~\eqref{HWG} in the next steps.

In our DNO-SM approach, we use Slater determinants as the intrinsic wave function ansatz. In Table~\ref{Exactsd}, we present the ground state energy of $^{24}$Mg in the sd valence space for both PAV and VAP schemes. 
The exact value is recovered with $16$ non-orthogonal Slater states in the VAP scheme compared to $975$ states in the PAV one where many-particle many-hole excitations on top of $16$ deformed Hartree-Fock states (column 2) are included. These results show that, even though in light systems such as $^{24}$Mg which is quite close to a realization of the SU(3) picture of a deformed rotor of Elliot, a proper consideration of many-particle many-hole ($N$p$N$h) excitations beyond the triaxial deformation $(\beta,\gamma)$ degree of freedom~\cite{RingSchuck1980} is necessary to describe fully the ground-state correlations. More spectacularly, the VAP approach using non-orthogonal Slater determinants capture perfectly these (tiny) correlations at the cost of a very few number of Slater states. In our recent work~\cite{DNOVAP2025}, we also show that these correlations emerge from the pairing component of the effective Hamiltonian. In summary, these results thus constitute a perfect illustration of the equivalence between the exact diagonalization and the variational principle $H\ket\Psi = E\ket\Psi \iff \delta\displaystyle\frac{\elmx{\Psi}{H}{\Psi}}{\overlap{\Psi}{\Psi}} = 0$. 
\begin{table}[H]
\centering
\scalebox{0.85}{
        \begin{tabular}{ccccc}
          \hline\hline %$E(0^+_{gs})$ (MeV)
 \multirow{2}{*}{$^{24}$Mg}         & \multicolumn{2}{c}{DNO-SM(PAV)} & \multirow{2}{*}{\textcolor{black}{DNO-SM(VAP)}} & \multirow{2}{*}{Exact SM} \\ \cline{2-3}
          & $(\beta,\gamma)$ & $(\beta,\gamma)$+\textcolor{blue}{$N$p$N$h} & \\ 
          \hline
%          \multirow{2}{*}{$^{20}$Ne} & $-40.35736$  & \textcolor{blue}{$-40.47233$}  &  \textcolor{red}{$-40.47231$}     &\textcolor{violet}{\textbf{$-40.47233$}} \\
%                                    & $7$ & $51$ & $3$   &  $640$ \\
Binding Energy  (MeV)         & $-86.73278$  & \textcolor{blue}{$-87.10428$}  &  \textcolor{red}{$-87.10405$}     &\textcolor{violet}{\textbf{$-87.10445$}} \\
 Number of basis states                                   & $16$ & $975$ & $16$   &  $28503$ \\
%          \multirow{2}{*}{$^{28}$Si} & $-135.21742$ & \textcolor{blue}{$-135.85891$} &  \textcolor{red}{$-135.86003$}     &\textcolor{violet}{\textbf{$-135.86073$}} \\
%                                    & $28$ & $4255$ & $45$   &  $93710$ \\
%          \multirow{2}{*}{$^{26}$Al} & $-$ & $-$ &  \textcolor{red}{$-105.74901$}     &\textcolor{violet}{\textbf{$-105.74934$}} \\
%                                    & $-$ & $-$ & $24$   &  $26914$ \\
          \hline\hline
      \end{tabular}}
      \caption{Comparison of the ground state binding energy (in MeV) between the DNO-SM(PAV) without (2nd column) and with $N$p$N$h excitations (3rd column), DNO-SM(VAP) (4th column) and exact Shell Model (Exact SM) calculations for $^{24}$Mg  using the USDB interaction.}
      \label{Exactsd}
\end{table}

\subsection{Pairing correlations in $^{48}$Cr with non-orthogonal Slater determinants}
\begin{figure}[H]
  \centering
  \includegraphics[scale=0.2]{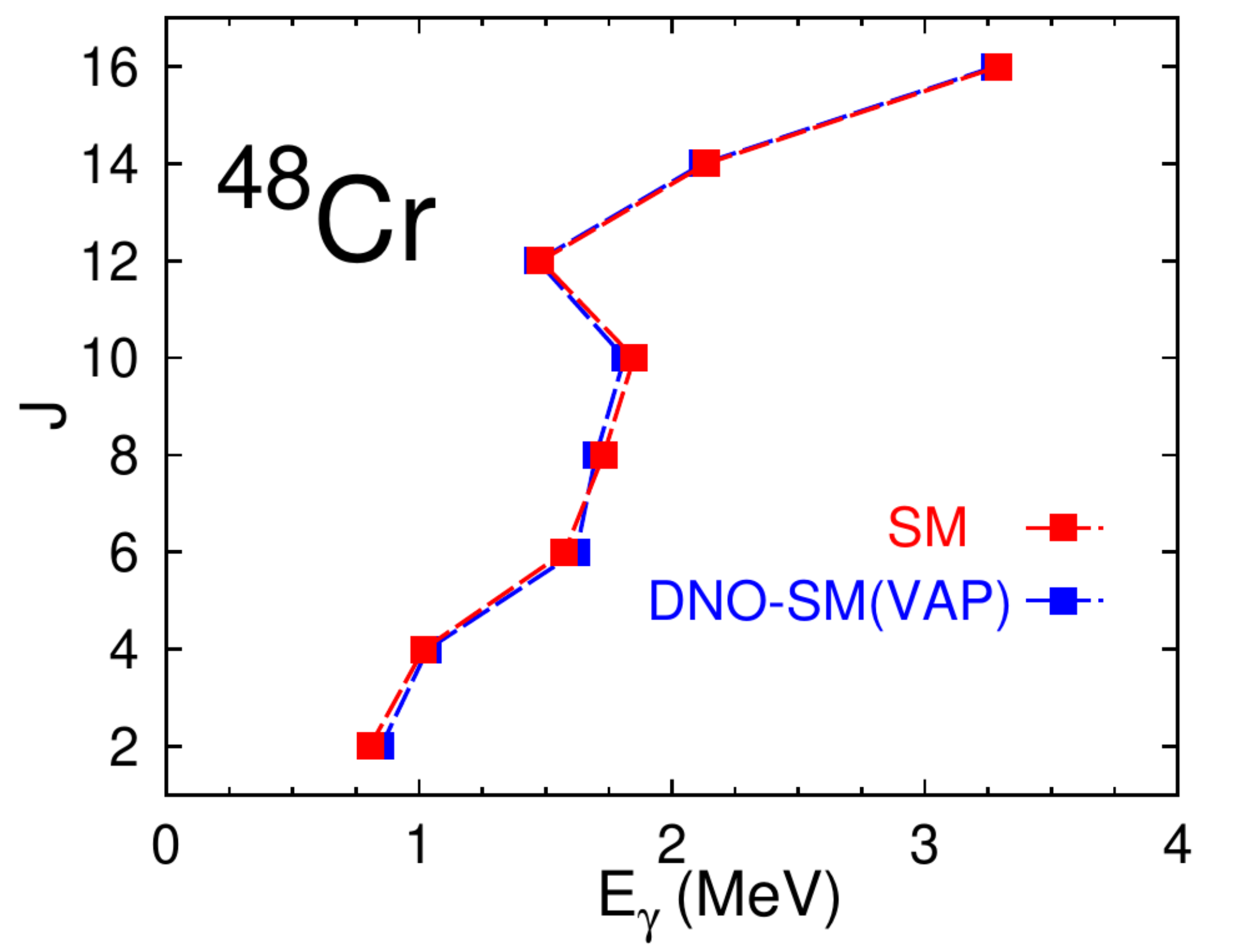}
  \caption{Comparison of the SM and DNO-SM calculations in the backbending pattern up to the angular momentum $J=16$. E$_\gamma$ is the energy of the corresponding gamma.}
  \label{Cr48backbend}
\end{figure}

To examine the performance of non-orthogonal Slater determinants in a more critical example, we consider the case of $^{48}$Cr where pairing correlations are known to produce the backbending phenomenon modifying the J(J+1) rotational pattern. The Kuo-Brown (KB3) effective interaction provides an excellent reproduction of this behavior~\cite{Caurier2005}. In Fig.~\ref{Cr48backbend}, the DNO-SM perfectly matches the shell-model results along the whole yrast band up to angular momentum $J=16$. In Table~\ref{Cr48gsTAB}, the corresponding absolute ground-state energy is recovered up to $3$ keV with $40$ VAP Slater states compared to the full pf diagonalization. 
\begin{table}[H]
    \centering
        \scalebox{1}{\begin{tabular}{cccc}
          \hline\hline
        $^{48}$Cr & DNO-SM$(\beta,\gamma)$ & DNO-SM(VAP) & Exact SM \\
          \hline
          Binding energy (MeV)   & $-31.873$ & \textcolor{blue}{$-32.950$} &  \textcolor{red}{$-32.953$} \\
          Number of basis states & $22$  & $40$  &  $1\:963\:461$  \\
          \hline\hline
      \end{tabular}}
    \caption{Comparison of the ground state binding energy (MeV) between the configuration mixing of deformed Hartree-Fock Slater states (DNO-SM$(\beta,\gamma)$), of non-orthogonal variation-after-projection Slater states (DNO-SM(VAP)) and the exact Shell Model (SM).}
    \label{Cr48gsTAB}
\end{table}

\subsection{Multiparticle-multihole mixing in $N\sim Z$ nuclei}
\begin{table}[H]
\centering
  \scalebox{0.88}{\begin{tabular}{cc*{5}c}
  \hline\hline
    &  \multicolumn{3}{c}{B(E2; $2_1^+ \rightarrow 0_1^+$) (e$^2$fm$^4$)}\\[1pt]
 nuclide   & DNO-SM & SM & Exp \\[-0pt]            
    \hline \\ [-1pt]
  $^{84}$Mo &    1512   &  -   & 1740$^{+580}_{-430}$ \\
  $^{86}$Mo &    893    &  731 &  707(71) \\  
  \hline
  \end{tabular}}
  \caption{Computed B(E2; $2_1^+ \to 0_1^+$) values. The values are obtained using  DNO shell-model calculations (DNO-SM) and  shell-model diagonalization (SM) and experimental values (Exp) for $^{84,86}$Mo.}  
  \label{Estimates}
\end{table}
The DNO Shell Model has been applied to the case of Molybdenum isotopes $^{84,86}$Mo to study the collapse of collectivity in these proton-rich nuclei recently observed~\cite{Recchia2025}. In Table~\ref{Estimates}, the Shell Model and DNO-SM computed E2 transitions are compared to the experimental data. The overall agreement is excellent suggesting a predominance of $8$p-$8$h intruder configurations in $^{84}$Mo and  our calculations point to a dramatic structure transition between $^{84}$Mo and $^{86}$Mo.
\begin{figure}[H]
\centering
\includegraphics[scale=0.3]{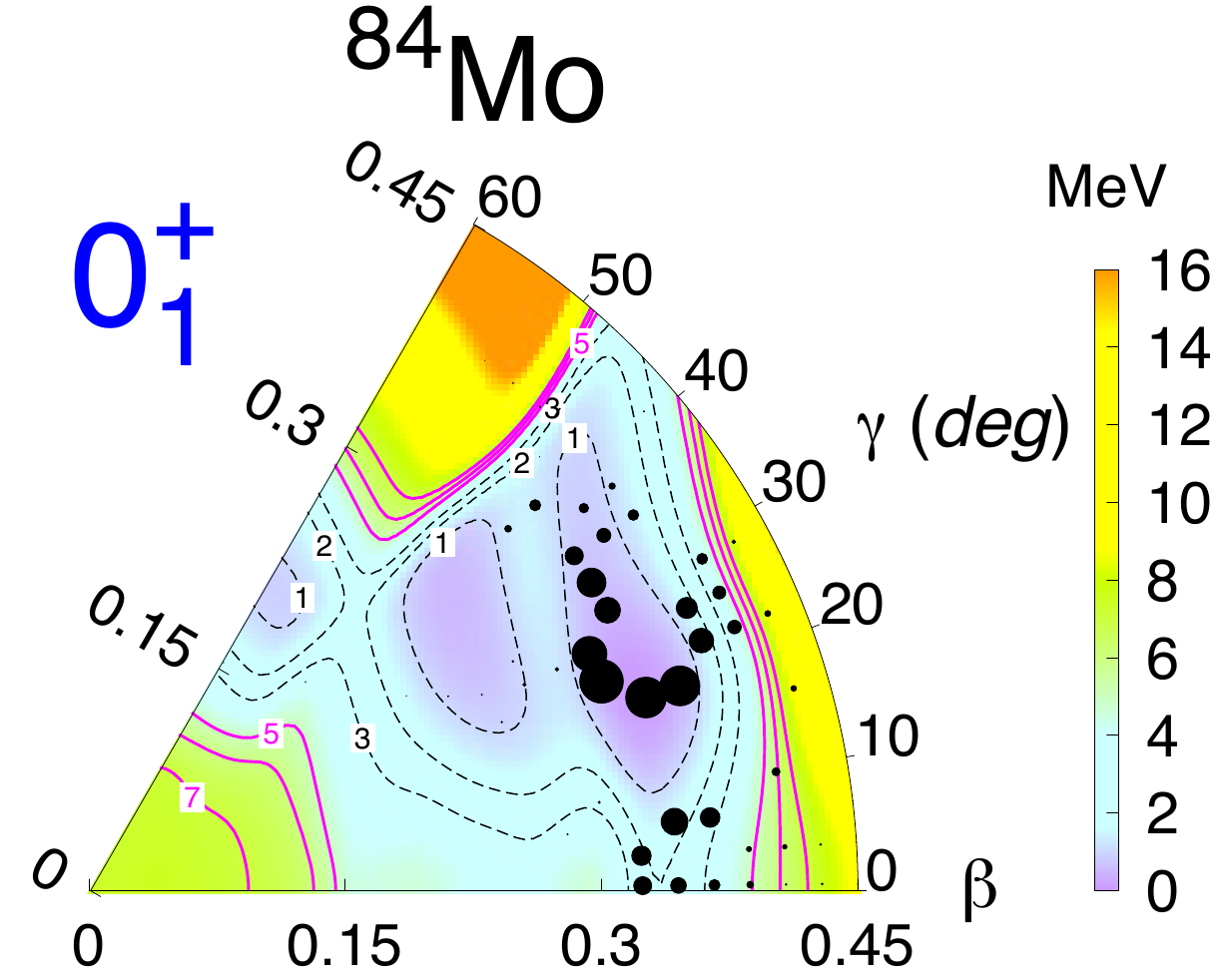}
\includegraphics[scale=0.3]{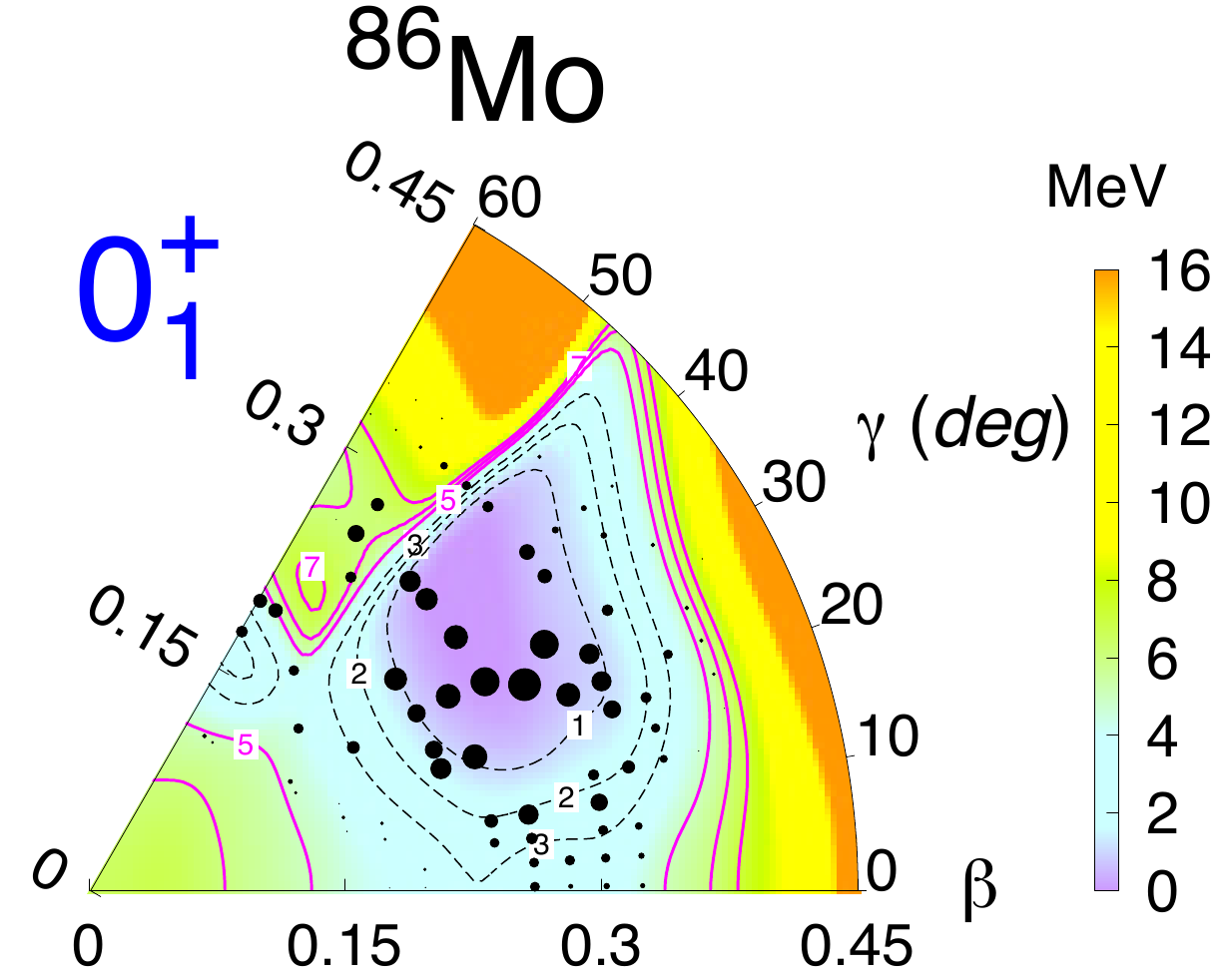}
\caption{Potential energy surfaces for $^{84,86}$Mo. The surfaces are obtained with the DNP-ZBM3 effective interaction. The area of the black circles represents  the probability of the $(\beta,\gamma)$ configurations in the ground state wave function.}
\label{PES}
\end{figure}
Indeed, as seen in Fig.~\ref{PES}, the potential energy surface landscape is quite flat and the $^{86}$Mo's ground-state wave function is an admixture of many configurations of equal amplitudes. However, in $^{84}$Mo, there are a few leading components of $8$p-$8$h nature whose contributions are the strongest. In~\cite{Recchia2025}, we argue that the origin of this behaviour can be traced back to the manifestation of three-body forces which should governs the evolution of g$9/2$-d$5/2$ valence orbitals along the $N\sim Z$ line. To illustrate this effect, calculations using a V$_{lowk}$-softened N$^3$LO two-nucleon ($2N$) interaction~\cite{N3LO-PhysRevC.68.041001} alone underestimate the $E2$ strength by more than one order of magnitude for both nuclei, producing a B(E2) of 174~e$^2$fm$^4$ for $^{84}$Mo and 0.63~e$^2$fm$^4$ for $^{86}$Mo. In summary, the observed abrupt transition in these nuclei suggests a termination of the island of deformation around $A\sim 80$.

\section{Spectroscopy of trans-actinide deformed nuclei: $^{254}$No}
\begin{figure}[H]
    %\centering
    %\vspace{-1.4em}
    \includegraphics[scale=0.35]{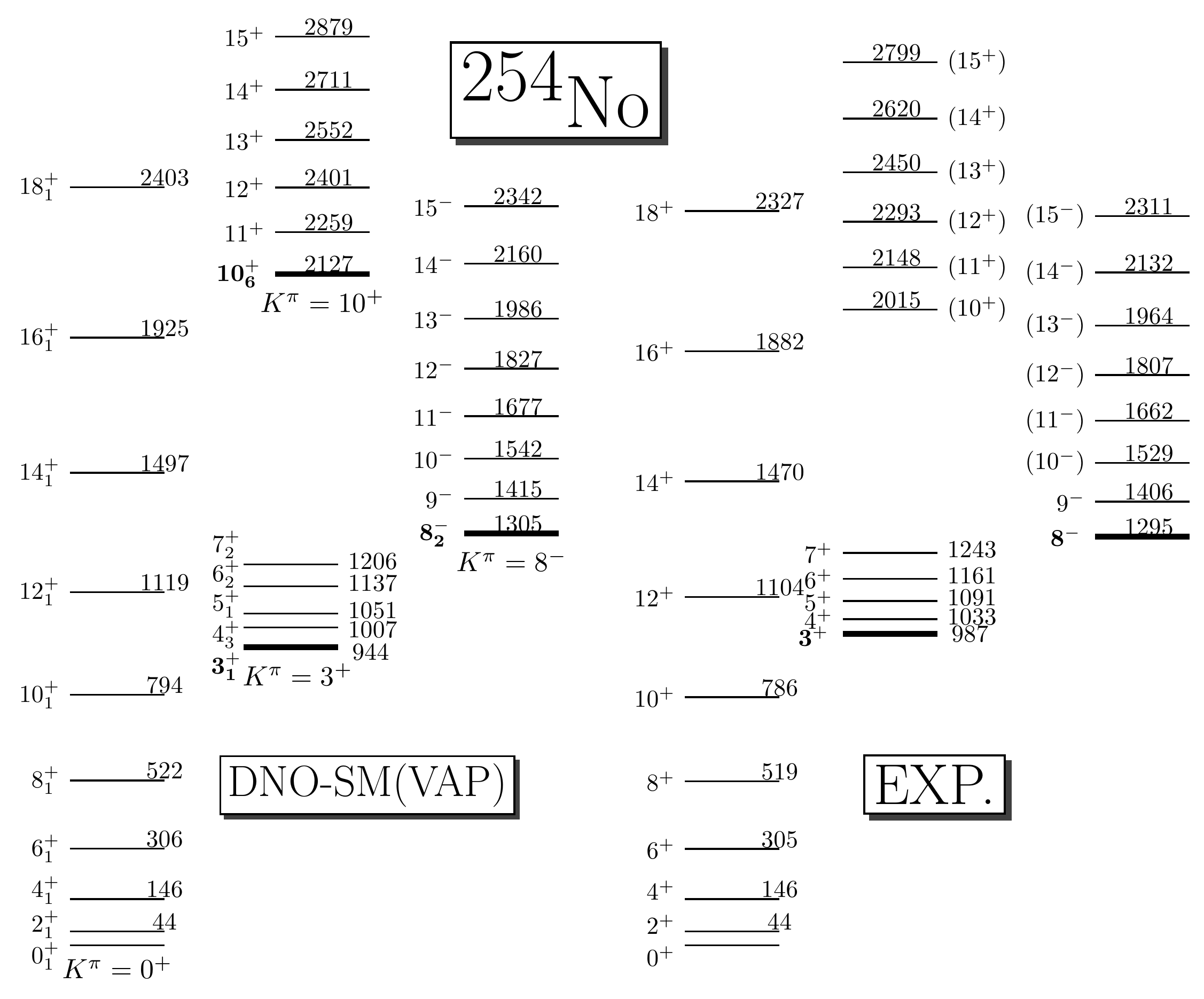}
    \caption{$^{254}$No spectrum from the DNO-SM(VAP) calculations compared to the experimental data (EXP). Energies are in keV.  For each band, $K$ denotes the total angular momentum projection quantum number.}
    \label{SM_254No}
\end{figure}

\begin{figure}[H]
  \centering
  \includegraphics[scale=0.4]{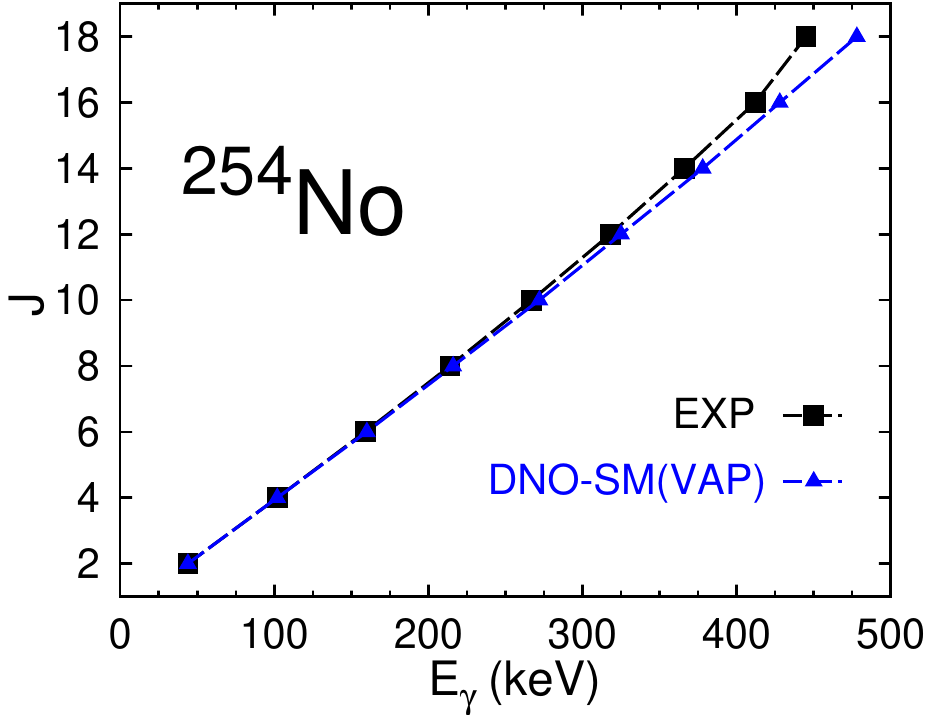}
  \includegraphics[scale=0.4]{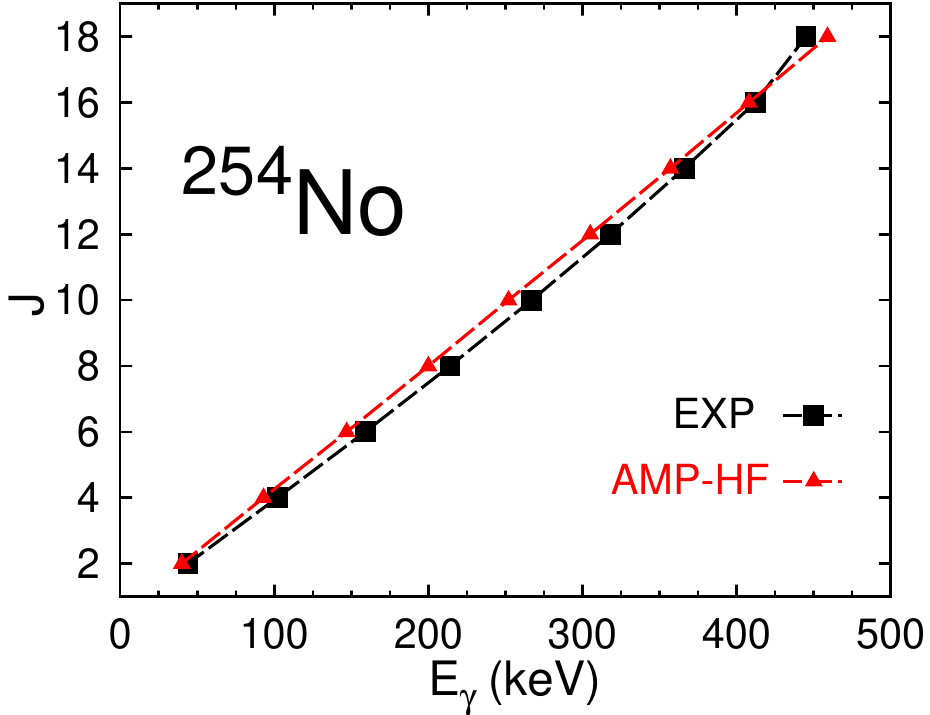}
  \caption{$\gamma$-excitation energies ($E_\gamma$) versus angular momentum ($J$) comparison with the experimental data for the yrast rotational band. The left panel corresponds to the full DNO-SM(VAP) calculation reported in Figure~\ref{SM_254No}. The right panel shows the yrast band with the angular momentum projection of the Hartree-Fock minimum (AMP-HF).}
  \label{rotor}
\end{figure}

Traditional shell-model diagonalization is so far believed to be intractable in the heavy actinides region because of the combinatorial dimensionality problem with respect to the system size. Although it is not the case from the point of view of variational approaches, a detailed Shell Model-like study in the region around the mass number $A\sim 250$ has never been performed before. With the recent DNO Shell Model under the variation-after-projection approach, dubbed as DNO-SM(VAP)~\cite{DNOVAP2025}, we present a first complete shell-model description~\cite{No254ARXIV2025} of one of the most studied nuclei in the $A\sim 250$ region, which marked the breakthrough in the experimental research of superheavy elements. As seen in Fig.~\ref{SM_254No}, our calculations reproduce strikingly all the known spectroscopy data: the yrast rotational band, the isomers $3^+$, $8^-$ and associated $K$-bands. The effective Kuo-Herling-based interaction is used for these predictions without adaptation to the current mass region. 
These results again show a tremendous predictive power of the Shell Model approach to obtain a quantitative description of low-energy properties of nuclei, as summarized in the Review of Modern Physics article by E. Caurier and collaborators~\cite{Caurier2005} about 20 years ago: \textit{"[...] two of the three “pillars” of the Shell Model: a good valence space and an effective interaction adapted to it. The third is a shell model code capable of coping with the secular problem."} To get a further insight into this successful description of $^{254}$No, let us recall here a fundamental feature of the classic rigid rotor whose spectrum follows $E_J = \displaystyle\frac{\hbar^2}{2\mathcal J_0} J(J+1)$ where $J$ is the spin and $\mathcal J_0$ the moment of inertia. As depicted in Fig.~\ref{rotor}, the experimental yrast band exhibits a quasi-perfect rotor behavior, which thus allows us to "assign" $^{254}$No as a well-deformed nucleus. 
In comparison with the case of $^{48}$Cr previously presented in Fig.~\ref{Cr48backbend}, we come to two interesting aspects:
\begin{itemize}
\item[i)] In heavy "well-deformed" nuclei, pairing correlations are weakened, due to strong deformation effects, to such an extent that the classic rotor model remains an excellent description of the energy spectrum; 
\item[ii)] The angular-momentum projected Hartree-Fock can be a very good approximation to generate rotational bands and essentially preserves the features of the classical rotor.
\end{itemize}
The second aspect ii) has been already known in the 1970s~\cite{Ripka1968} where the Hartree-Fock approximation is shown to capture essentially the deformation component of the effective interaction. Indeed, this aspect is reflected in the right panel of Figure~\ref{rotor} with the Hartree-Fock minimum projected onto good angular momentum. From these considerations, while the power of the VAP combined with Slater determinants is now understandable and indeed a very welcome feature for a microscopic theoretical study of such heavy nuclei, a deep insight into the physical description should eventually be, to quote Andr\'e Zuker, adhered to the nuclear monopole-multipole folklore: \textit{"Pairing plus Quadrupole propose, Monopole disposes."}~\cite{Zuker1993CRN,Zuker1996_PhysRevC.54.1641}. 

\section{Conclusions}
In summary, with the newly developed DNO Shell Model, we have discussed some of its recent applications to very challenging cases in nuclear structure. After examining the capacity of the variation after projection combined with Slater determinants, in the light of the Broeckhove-Deumens theorem, we have shown that it is possible to solve the secular shell-model problem in a very efficient way and to fully capture ground-state correlations of pairing nature~\cite{DNOVAP2025}. Applications of the DNO Shell Model to the case of Molybdenum isotopes has identified a complex case of structure transition in the proton-rich side along the $N\sim Z$ line~\cite{Recchia2025}. A first detailed comparison of low-lying spectroscopy in the superheavy $^{254}$No has been performed. The theoretical findings are in remarkable agreement with all the known data so far. We believe that all these results will trigger more theoretical and experimental investigations, in particular in the superheavy region, and the future would hold great promise for the Shell Model as an unified view in nuclear structure studies.  

\bibliography{main}

\begin{thebibliography}{16}

\bibitem{Caurier2005}
E.~Caurier, G.~Mart\'{\i}nez-Pinedo, F.~Nowacki, A.~Poves, A.P. Zuker, The
  shell model as a unified view of nuclear structure, Rev. Mod. Phys.
  \textbf{77}, 427 (2005). \doiwoc{10.1103/RevModPhys.77.427}

\bibitem{BigStickJOHNSON20132761}
C.W. Johnson, W.E. Ormand, P.G. Krastev, Factorization in large-scale many-body
  calculations, Comp. Phys. Comm. \textbf{184}, 2761 (2013).
  \doiwoc{10.1016/j.cpc.2013.07.022}

\bibitem{NuShellXBROWN2014}
B.~Brown, W.~Rae, The shell-model code nushellx@msu, Nucl. Data Sheets
  \textbf{120}, 115 (2014). \doiwoc{10.1016/j.nds.2014.07.022}

\bibitem{NOWACKI2021}
F.~Nowacki, A.~Obertelli, A.~Poves, The neutron-rich edge of the nuclear
  landscape: Experiment and theory, Prog. Part. Nucl. Phys. \textbf{120},
  103866 (2021). \doiwoc{10.1016/j.ppnp.2021.103866}

\bibitem{Schmid1984A_PhysRevC.29.291}
K.W. Schmid, F.~Gr\"ummer, A.~Faessler, Nuclear structure theory in spin- and
  number-conserving quasiparticle configuration spaces: General formalism,
  Phys. Rev. C \textbf{29}, 291 (1984). \doiwoc{10.1103/PhysRevC.29.291}

\bibitem{Schmid2004}
K.~Schmid, On the use of general symmetry-projected hartree–fock–bogoliubov
  configurations in variational approaches to the nuclear many-body problem,
  Prog. Part. Nucl. Phys. \textbf{52}, 565 (2004).
  \doiwoc{10.1016/j.ppnp.2004.02.001}

\bibitem{DNO2022_PhysRevC.105.054314}
D.D. Dao, F.~Nowacki, Nuclear structure within a discrete nonorthogonal shell
  model approach: New frontiers, Phys. Rev. C \textbf{105}, 054314 (2022).
  \doiwoc{10.1103/PhysRevC.105.054314}

\bibitem{DNOVAP2025}
D.~D.~Dao, F.~Nowacki, Exact solutions of the nuclear shell-model secular
  problem: Discrete non-orthogonal shell model within a variation after
  projection approach, arXiv:2507.09073  (2025).
  \doiwoc{10.48550/arXiv.2507.09073}

\bibitem{Deumens1979}
J.~Broeckhove, E.~Deumens, A mathematical foundation for discretisation
  techniques in the generator coordinate method, Z. Phys. A \textbf{292},
  243–247 (1979). \doiwoc{10.1007/BF01547468}

\bibitem{Recchia2025}
F.~Recchia et~al., {Abrupt Structural Transition in Proton-Rich Molybdenum
  Isotopes unveils an Isospin-Symmetric Island of Inversion (accepted, in
  production)}, Nature Comm.  (2025).

\bibitem{No254ARXIV2025}
D.~D.~Dao, F.~Nowacki, First complete description of low-lying spectroscopy in
  $^{254}$\rm{No}, arXiv:2409.08210  (2025). \doiwoc{10.48550/arXiv.2409.08210}

\bibitem{RingSchuck1980}
P.~Ring, P.~Schuck, The nuclear many-body problem (Springer-Verlag, 1980)

\bibitem{N3LO-PhysRevC.68.041001}
D.R. Entem, R.~Machleidt, Accurate charge-dependent nucleon-nucleon potential
  at fourth order of chiral perturbation theory, Phys. Rev. C \textbf{68},
  041001 (2003).

\bibitem{Ripka1968}
G.~Ripka, The Hartree-Fock Theory of Deformed Light Nuclei (Springer US, 1968),
  pp. 183--259

\bibitem{Zuker1993CRN}
A.P. Zuker, Random and coherent behaviour in nuclear hamiltonians, Centre de
  Recherches Nucl\'eaires de Strasbourg, Internal Report 29  (1993).

\bibitem{Zuker1996_PhysRevC.54.1641}
M.~Dufour, A.P. Zuker, Realistic collective nuclear hamiltonian, Phys. Rev. C
  \textbf{54}, 1641 (1996). \doiwoc{10.1103/PhysRevC.54.1641}

\end{thebibliography}

\end{document}